\documentclass[8.5pt,twoside,twocolumn]{article}
\usepackage[super,sort&compress,comma]{natbib} 
\usepackage{mhchem}
\usepackage{times}
\usepackage{sectsty}
\usepackage{balance} 

\usepackage{graphicx} 
\usepackage{lastpage}
\usepackage[format=plain,justification=raggedright,singlelinecheck=false,font=small,labelfont=bf,labelsep=space]{caption} 
\usepackage{fancyhdr}
\begin{document}

\thispagestyle{plain}
\fancypagestyle{plain}{
\fancyhead[L]{\includegraphics[height=8pt]{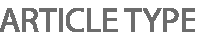}}
\fancyhead[C]{\hspace{-1cm}\includegraphics[height=20pt]{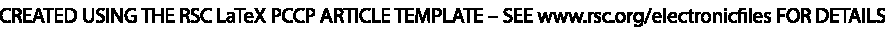}}
\fancyhead[R]{\includegraphics[height=10pt]{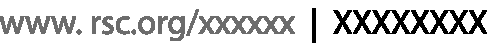}\vspace{-0.2cm}}
\renewcommand{\headrulewidth}{1pt}}
\renewcommand{\thefootnote}{\fnsymbol{footnote}}
\renewcommand\footnoterule{\vspace*{1pt}%
\hrule width 3.4in height 0.4pt \vspace*{5pt}} 
\setcounter{secnumdepth}{5}

\makeatletter 
\def\subsubsection{\@startsection{subsubsection}{3}{10pt}{-1.25ex plus -1ex minus -.1ex}{0ex plus 0ex}{\normalsize\bf}} 
\def\paragraph{\@startsection{paragraph}{4}{10pt}{-1.25ex plus -1ex minus -.1ex}{0ex plus 0ex}{\normalsize\textit}} 
\renewcommand\@biblabel[1]{#1}            
\renewcommand\@makefntext[1]%
{\noindent\makebox[0pt][r]{\@thefnmark\,}#1}
\makeatother 
\renewcommand{\figurename}{\small{Fig.}~}
\sectionfont{\large}
\subsectionfont{\normalsize} 

\fancyfoot{}
\fancyfoot[LO,RE]{\vspace{-7pt}\includegraphics[height=9pt]{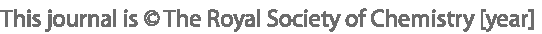}}
\fancyfoot[CO]{\vspace{-7.2pt}\hspace{12.2cm}\includegraphics{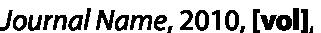}}
\fancyfoot[CE]{\vspace{-7.5pt}\hspace{-13.5cm}\includegraphics{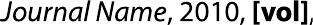}}
\fancyfoot[RO]{\footnotesize{\sffamily{1--\pageref{LastPage} ~\textbar  \hspace{2pt}\thepage}}}
\fancyfoot[LE]{\footnotesize{\sffamily{\thepage~\textbar\hspace{3.45cm} 1--\pageref{LastPage}}}}
\fancyhead{}
\renewcommand{\headrulewidth}{1pt} 
\renewcommand{\footrulewidth}{1pt}
\setlength{\arrayrulewidth}{1pt}
\setlength{\columnsep}{6.5mm}
\setlength\bibsep{1pt}

\twocolumn[
  \begin{@twocolumnfalse}
\noindent\LARGE{\textbf{Absorbing Phase Transitions and Dynamic Freezing in Running Active Matter Systems }}
\vspace{0.6cm}

\noindent\large{\textbf{Charles Reichhardt and
Cynthia J. Olson Reichhardt,$^{\ast}$}}\vspace{0.5cm}

\noindent\textit{\small{\textbf{Received Xth XXXXXXXXXX 20XX, Accepted Xth XXXXXXXXX 20XX\newline
First published on the web Xth XXXXXXXXXX 200X}}}

\noindent \textbf{\small{DOI: 10.1039/b000000x}}
\vspace{0.6cm}

\noindent \normalsize{
We examine a two-dimensional system of sterically 
repulsive interacting disks where each particle 
runs in a random direction. 
This system is equivalent to a 
run-and-tumble dynamics system in the limit where 
the run time is infinite.
At low densities, we find a
strongly fluctuating state composed of transient clusters. 
Above a critical density that is well below the density 
at which non-active particles would crystallize, 
the system can organize into a 
drifting quiescent or frozen state 
where the fluctuations are lost and large crystallites form
surrounded by a small density of individual particles. 
Although all the particles are still moving, their paths
form closed orbits. The average transient time to organize into  the
quiescent state diverges as a power law upon
approaching the critical density from above.
We compare  
our results to the random organization 
observed for periodically sheared systems that can 
undergo an absorbing transition  
from a fluctuating state to a dynamical non-fluctuating state. 
In the random organization studies, 
the system organizes to a state 
in which the particles no longer interact; in
contrast, we find that the randomly running active matter organizes 
to a strongly interacting dynamically jammed state. 
We show that the transition to the frozen state is robust 
against a certain range of stochastic fluctuations.    
We also examine the effects of adding a small number of pinned 
particles to the system and find that the transition to the frozen
state shifts to significantly lower densities and arises via the
nucleation of 
faceted crystals centered at the obstacles.
}
\vspace{0.5cm}
 \end{@twocolumnfalse}
  ]

\section{Introduction}

\footnotetext{\textit{Theoretical Division, Los Alamos National Laboratory, Los Alamos, New Mexico 87545, USA. Fax: 1 505 606 0917; Tel: 1 505 665 1134; E-mail: cjrx@lanl.gov}}

Nonequilibrium collections of interacting particles  
often form strongly fluctuating or turbulent states and 
can exhibit nonequilibrium phase transitions 
between distinct dynamical regimes \cite{1,2,3}. Examples
of such systems include sheared 
colloids \cite{4,5,6,7}, 
turbulent liquid crystals \cite{8}, 
granular matter \cite{9,10}, driven vortex matter in 
type-II superconductors \cite{11,12}, 
and active matter or self-driven particle assemblies \cite{13,14,15,17}. 
In certain cases a nonequilibrium transition can be 
identified to occur between a fluctuating state and a
quiescent state in which fluctuations are almost completely
absent but the system is still out of equilibrium or even in motion; here,
the system can become dynamically trapped into an absorbing phase \cite{1,2}. 
The absence of fluctuations in the absorbing or quiescent state makes it
impossible for the system to escape from this state.

Recently, periodically sheared dilute colloidal suspensions were shown to be
an outstanding example of a system exhibiting 
the hallmark features of a transition
into an absorbing 
phase \cite{4,5,6,7}.
Thermal fluctuations are negligible in this system, so the dynamics is
governed by collisions between the particles. 
Both numerical
and experimental studies compared the position of the colloidal particles
at the beginning and end of each shear cycle.
This protocol permits the identification of
two distinct dynamical regimes. 
In the fluctuating or irreversible state, 
the particles are at different locations at the beginning and end
of each cycle, and perform an anisotropic random walk over the course of many
cycles.
In the reversible state, particles return to the same position after each
cycle and
fluctuations vanish when the particles organize
into a configuration that prevents particle-particle collisions from
occurring \cite{4,5}.
For fixed particle density, the reversible state appears below a 
critical oscillatory shear amplitude $\gamma_{c}$, 
while at high shear amplitudes the system remains in the fluctuating state. 
When the periodic shear is first applied, the system always starts in
an irreversible fluctuating state, but after a number of cycles it either
organizes into a steady irreversible state or reaches a reversible
quiescent state.
As $\gamma_{c}$ is approached, the  
number of cycles or the total time $\tau$ required for the system to 
reach one of these states diverges as a power law 
$\tau \propto |\gamma -\gamma_{c}|^{-\nu_{||}}$, 
with $\nu_{||} \approx 1.35$ in two-dimensional (2D) simulations 
and $\nu_{||} \approx 1.5$ in three-dimensional (3D) experiments \cite{5}.
Major classes of absorbing phase transitions include directed percolation 
\cite{1,18,19}
and conserved directed percolation \cite{2,19},
which have predicted exponents in 2D of $\nu_{||}=1.295$ and
$\nu_{||}=1.225$, respectively.

In the colloidal system
the absorbing state is characterized by the organization of the particles
into a random pattern in which they do not interact with each other,
leading the transition to be termed ``random organization'' \cite{5,19}.
A similar periodic drive protocol has been used to analyze other nonequilibrium 
systems, such as periodically driven superconducting vortices \cite{11,12} and
granular media \cite{19,20,21},
by identifying the transient time required to transition from
irreversible to reversible flow.
Other studies have shown that a transition from an irreversible to a
reversible state can occur even when the reversible state is strongly
interacting, such as in jammed solids
\cite{22,23,24,25}, and in some cases this transition is associated with
a power law divergence of a time scale, indicative of a critical
point \cite{22,23,25}. 

Another type of nonequilibrium systems is self-driven particles or 
active matter \cite{13,17,26},
which include
swimming bacteria undergoing run-and-tumble dynamics \cite{27,28,29,30,30.N}, 
flocking particles \cite{31,32}, pedestrian motion \cite{33}, 
and self-driven colloidal particles \cite{34,35,36,37,38,39,40}.
One class of active matter is run-and-tumble 
self-driven particles, which move or run
in a randomly chosen direction for a period of time before 
undergoing a tumbling event and then moving in a freshly chosen
random direction.  Run-and-tumble dynamics have been used to model
systems such as
swimming bacteria \cite{27,28,29,30,38}.
Another active matter class is active Brownian particles  
such as Janus colloids.  Here each particle is self-driven but its
direction of motion slowly diffuses
\cite{34,36,37,38,39}.
When the active particles have additional steric interactions, 
a transition can occur 
from a uniform liquid state to 
a phase-separated state consisting of high density clusters separated by a low 
density dilute gas \cite{30,37,38,39,40,41,42,43,44,45,46,lev14,pohl14}. 
The high density state 
can be characterized as having glass \cite{43,46} or 
crystalline \cite{37,39,42,44,45} order. 
The 
cluster structures are only transient and exhibit fluctuations in which
the clusters can break up and reform over time, 
as observed in recent experiments 
where the clusters were described as ``living crystals'' \cite{37,39}.
When the particle density is increased, the system eventually enters 
a dense jammed state in which slow rearrangements of the particles can still
occur due to their activity
\cite{41,47}. 

Here we investigate whether active matter systems can exhibit 
absorbing transitions
similar to the random organization from a strongly fluctuating state 
to a dynamically frozen steady state. 
Active matter systems 
such as run-and-tumble and active Brownian particles  
are usually modeled by including a stochastic term either in
the randomization of the running direction during a tumble or
in the slow diffusion of the running direction, respectively.
Deterministic dynamics can occur in certain limits, such as 
infinite run lengths in the run-and-tumble systems, or in the
absence of a diffusive term in the active Brownian particles.
In these deterministic limits, each particle moves in a fixed direction  
but can still interact with other particles to create a fluctuating
state.
When the dynamics is free of stochastic fluctuations,
as in recent numerical studies
of binary rotating cross-shaped particles, 
it was shown that after some time particles with different rotation
directions could organize into a phase-separated state \cite{48}.
This system is similar to the random organization studies
in that the particles are initially in a fluctuating state and after some
time either settle into a dynamically frozen state or remain in a
continuously fluctuating state \cite{5}.
Other studies of growing actin filaments
found evidence of 
organization into a quiescent absorbing state
consisting of spiral patterns \cite{14,15}.

We consider run-and-tumble sterically repulsive disks 
in the limit where the particles
have an infinite run time. 
The initial positions of the particles and the run directions are 
selected randomly. 
At low density
the system forms a strongly fluctuating state 
containing short-lived clusters; however, above a critical
density $\phi_{c}$ 
the system can organize into a non-fluctuating or 
quiescent state comprised of a single 
drifting crystalline cluster surrounded 
by a number
of individual particles that are traveling in closed orbits. 
By starting from the high density limit and decreasing 
the density to $\phi_{c}$, we find
that the time required for the system to organize to a
frozen state increases 
with decreasing $\phi$ and diverges as a power law at $\phi_c$
with $\nu_{||} = 1.21 \pm 0.3$, 
consistent with the 
exponents found in the random organization system and 
conserved directed percolation. 
We show that the transition is robust against the addition of
some thermal noise;
however, 
at high enough noise it is no longer possible for a completely
dynamically frozen state to form.
When we add a small number of pinned particles, 
we find that the system can organize
to a dynamically frozen pinned state at 
densities well below the obstacle-free $\phi_{c}$.
The pinned frozen states contain faceted hexagonal crystals that  
are centered at the obstacle sites.

\section{Simulation and System}   
We consider a 2D system 
of size $L \times L$ with periodic boundary conditions in the 
$x$ and $y$-directions. 
The sample contains $N$ active disks 
with radius $R_{d}$ that interact
with each other via
a steric harmonic repulsion. 
The disk density $\phi=\pi NR_d^2/L^2$ is
the total fraction of the sample area covered by the disks. 
A sample filled with inactive disks crystallizes into a triangular
lattice at a density of $\phi \approx 0.9$.
The motion of an individual disk $i$ 
located at ${\bf r}_i$ is obtained by integrating
the overdamped equations of motion,
\begin{equation}  
\eta \frac{d {\bf r}_{i}}{dt} = 
 {\bf F}_i^{m} + {\bf F}^{s}_i .
\end{equation} 
Here $\eta$ is the damping constant, 
${\bf F}^{m}_i=F_d{\bf \hat m}_i$ is the motor force, and
${\bf F}^{s}$ describes the particle-particle 
interactions. 
The motor force consists of a fixed force $F_d$ applied in a
fixed direction ${\bf \hat m}_i$ that is randomly selected for each disk
at the beginning of the simulation and then never changed.
For run-and-tumble dynamics this 
corresponds to an infinite run length, while for active Brownian particles
it would correspond to zero rotational diffusion of the particle motion.
The steric disk-disk repulsion is 
${\bf F}^s_{i} = \sum^{N}_{i\ne j}k(2R_{d} - |{\bf r}_{ij}|)\Theta(2R_{d} - |{\bf r}_{ij}|){\hat {\bf r}}_{ij}$,
where ${\bf r}_{ij}={\bf r}_i-{\bf r}_j$
and ${\bf \hat r}_{ij}={\bf r}_{ij}/|{\bf r}_{ij}|$.
We set $R_{d} = 0.5$ and select parameters $k=20$ and $F_d=0.5$ such that 
disks cannot squeeze past a jammed configuration due to the driving force.
Pinned particles or obstacles are introduced in the form of 
$N_p$ immobile disks that 
have the same steric repulsive interactions as the mobile disks.
Their density is given by $\phi^p=N_p\pi R^2_d/L^2.$

\section{Results}

\begin{figure}
\centering
\includegraphics[width=0.5\textwidth]{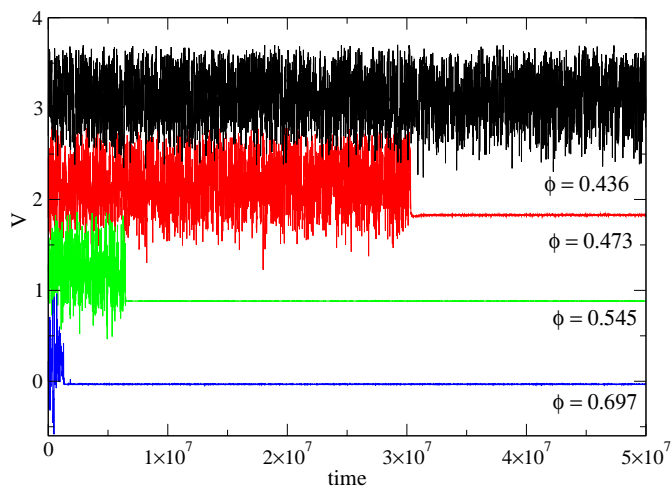}
\caption{
The velocity fluctuations $V(t)$ of 
a single probe particle driven in the $x$ direction for 
disk densities $\phi = 0.436,$ 0.473, 0.545, and $0.697$, 
from top to bottom. The curves have been vertically shifted for clarity.
For $\phi=0.473$ and above, there is a transition
from a strongly fluctuating state to a non-fluctuating state after an
average time that decreases with increasing $\phi$.
Below a critical $\phi_{c}$, the system 
remains in the fluctuating state.     
}
\label{fig:1}
\end{figure}

To characterize the activity in the system, 
we measure the velocity fluctuations
$V(t)$ at varied $\phi$
of a single probe particle $p$ moving in the positive $x$-direction,
where $V(t)={\dot x}_p(t)$ and ${\bf \hat m}_p={\bf \hat x}.$
Figure~\ref{fig:1} shows a series of $V(t)$ curves taken at
$\phi = 0.436$, 0.473, 0.545, and $0.697$.
At $\phi = 0.697$, the system starts in a fluctuating state with
transient clusters or living crystals, 
but after $1.5\times 10^6$    
simulation time steps the fluctuations almost completely vanish and 
$V \approx 0$.
The other particles in the system exhibit the same behavior as the
probe particle, and have the same transition to a nonfluctuating velocity
occurring at the same time.
The time required to reach the non-fluctuating state is labeled $\tau$. 
For $\phi  = 0.545$ we find the same
transition to a non-fluctuating state 
but $\tau$ is larger, and as $\phi$ is further decreased $\tau$ continues
to increase as shown for
$\phi = 0.473$.  For $\phi < 0.46$, the system never reaches a 
non-fluctuating state
even for extremely long simulation times.    

\begin{figure}
\centering
\includegraphics[width=0.5\textwidth]{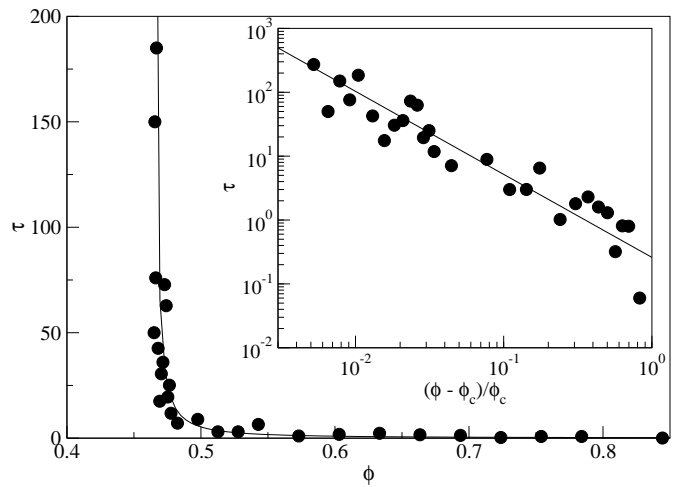}
\caption{
The transient time $\tau$ to reach a dynamically frozen state vs $\phi$. 
The solid line is a fit to 
$\tau = (\phi -\phi_{c})^{-\nu_{||}}$, with $\phi_{c} = 0.462$ and 
$\nu_{||} = 1.21$.
The $\tau$ axis is divided by $10^6$ simulation time steps.
Inset: The same data on a log-log plot showing the power law fit.  
}
\label{fig:2}
\end{figure}

In any individual simulation at a given value of $\phi$ there is some variation
in the value of $\tau$ when different random initial conditions are chosen;
however, the average value of $\tau$ robustly increases with decreasing
$\phi$ and we find that $\tau$ diverges at a critical density $\phi_c$.
In Fig.~\ref{fig:2} we plot $\tau$ versus $\phi$, 
where the solid line is a fit 
to the form $\tau = (\phi -\phi_{c})^{-\nu_{||}}$  
using $\phi_{c} = 0.462$ and $\nu_{||} = 1.21$. 
The inset of Fig.~\ref{fig:2} shows
a log-log plot of $\tau$ 
versus $(\phi -\phi_{c})/\phi_{c}$ fit using $\nu_{||} = 1.21 \pm 0.06$.    
The diverging time scale for the transition 
from a fluctuating state to a dynamically frozen state 
is very similar to what is observed
in the periodically sheared colloidal system, 
where a diverging time scale appears at a critical shear amplitude
with exponent $\nu_{||}=1.33$ measured in 2D colloidal simulations \cite{5}.
Other simulations of a diverging time scale at the transition to
a dynamically frozen state for a periodically sheared disk system give
$\nu_{||} = 1.3$ \cite{19}. 
These exponents are close to the values $\nu_{||}=1.295$ and
$\nu_{||}=1.225$ expected for 2D directed percolation (DP) \cite{18,19} and
2D conserved directed percolation (CDP) \cite{2,19}, respectively.
Our regression fit gives 
a value of $\nu_{||}$ closer to the expected value for
2D CDP;
however, our results are not accurate enough to positively 
distinguish between DP and CDP.   We note that in our
system the particle number is conserved.

\begin{figure}
\centering
\includegraphics[width=0.5\textwidth]{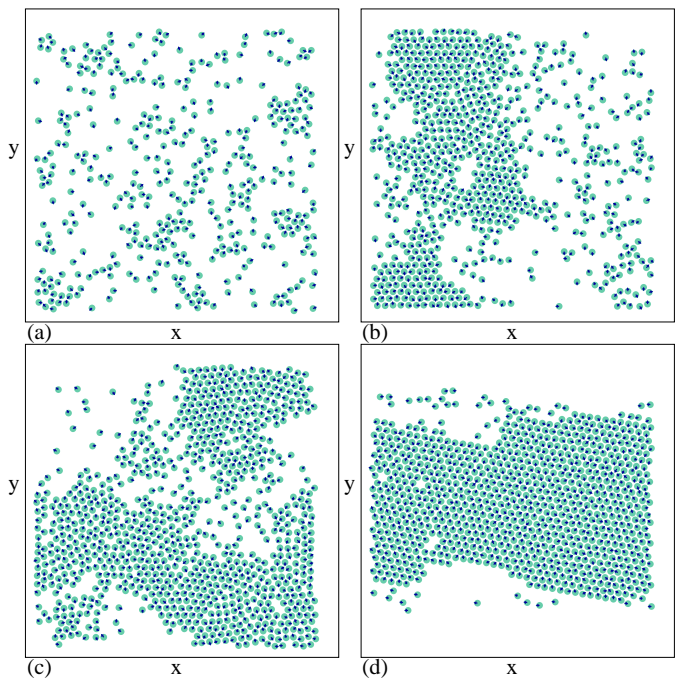}
\caption{
Snapshots of disk positions 
where the arrows indicate 
the direction of the motor force for each disk. 
(a) The fluctuating state at $\phi = 0.242$ where only small transient 
clusters form.
(b) The fluctuating state at $\phi = 0.436$ 
where a living crystal cluster forms.
(c) The transient fluctuating state at $\phi = 0.545$. 
(d) The dynamically frozen steady state at $\phi = 0.545$,    
where a single large crystal forms and drifts through the system. 
}
\label{fig:3}
\end{figure}

The non-fluctuating state in the random organization system consists of
particles that are moving but that have organized into a dynamic configuration
in which they no longer collide with one another \cite{5}.
In contrast, we find that the active matter
system organizes into  
a non-fluctuating cluster where most of the particles are in 
contact with each other. 
We note that the cluster state is still dynamical since it continues to 
drift through the system in a direction determined by the net sum of the
motor forces of all the particles contained in the cluster.
In Fig.~\ref{fig:3} we illustrate the disk positions
along with the direction of the motor force for each disk. 
Figure \ref{fig:3}(a) shows a fluctuating state 
for $\phi < \phi_{c}$ at $\phi =0.242$. 
At this density the system 
forms transient clusters that change rapidly.
At $\phi=0.436$ in Fig.~\ref{fig:3}(b), below $\phi_c$, the system phase
separates
into living crystal clumps 
with triangular ordering which break apart over time, as observed in
other active matter studies
\cite{37,39,42,43,44,45,46}. 

In Fig.~\ref{fig:3}(c) we illustrate
the transient active fluctuating state 
at $\phi = 0.545$ which is above $\phi_{c}$, while in Fig.~\ref{fig:3}(d)
we show the dynamically frozen steady state that forms at later times.
In the frozen state, the disks form a
single crystalline structure
that does not change, and the particles within the
cluster maintain the same neighbors over time.
The entire cluster slowly drifts across the sample as a 
function of time, while the disks that 
are not part of the 
cluster follow closed orbits which repeat over 
time due to the periodic boundary conditions. 
The system can be regarded as organizing not into a random state but rather
into a jammed non-fluctuating state.
Due to the absence of fluctuations, the particles in the cluster remain
trapped in the cluster.
In the fluctuating state the disks can explore various 
configurations and when they encounter
a configuration in which all fluctuations are suppressed, they become
trapped in this configuration 
even though all of the particles are still subject to
a motor force and the entire cluster assembly is moving as a rigid object.
The number of configurations 
that produce frozen states increases with increasing $\phi$,  
so the average time $\tau$ required for the particles to 
find a dynamically frozen state decreases 
with increasing $\tau$ as shown in Figs.~\ref{fig:1},\ref{fig:2}. 

\begin{figure}
\centering
\includegraphics[width=0.5\textwidth]{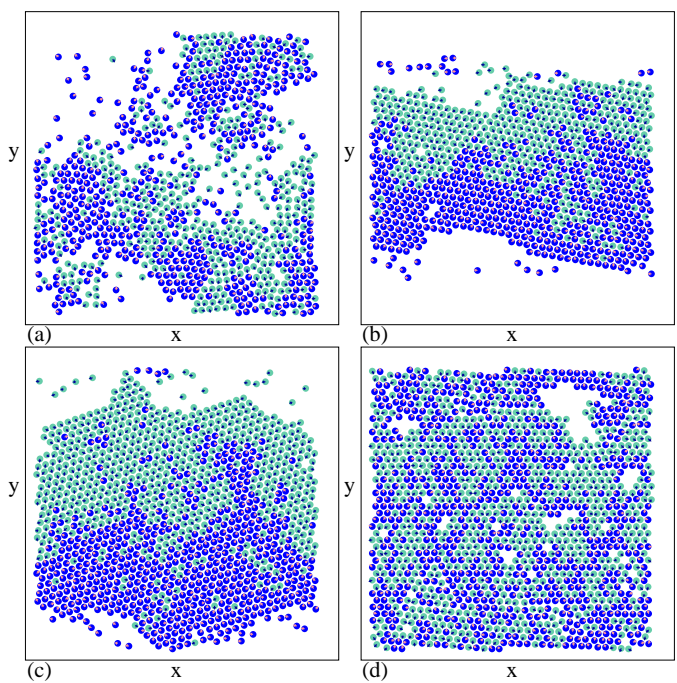}
\caption{
Snapshots  of the disk positions 
where the arrows indicate the direction 
of the motor force for each disk.  Dark 
particles 
with light arrows 
have
a net motion in the positive $y$ direction; 
light 
particles 
with dark arrows
have a net motion in the negative $y$ direction.
(a) The transient fluctuating state at $\phi = 0.545$. 
(b) The dynamically frozen steady state at $\phi = 0.545$ 
where a single large cluster forms. 
Particles in the lower half of the cluster are generally
moving in the positive $y$ direction, 
while particles in the upper half of the cluster are generally
moving in the negative $y$ direction, creating a
jammed state.
(c) The dynamically frozen steady state at $\phi = 0.73$. 
(d) The dynamically frozen steady state at $\phi = 0.848$,    
where a single large crystalline structure forms 
that drifts through the system. 
}
\label{fig:4}
\end{figure}

\begin{figure}
\centering
\includegraphics[width=0.5\textwidth]{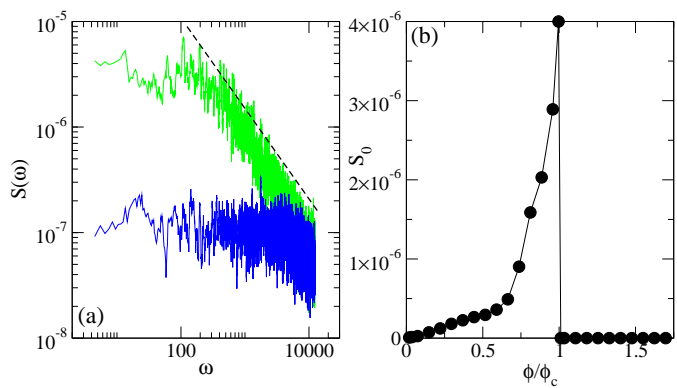}
\caption{
(a) Power spectrum $S(\omega)$ of the time series 
$V(t)$ of the velocity fluctuations for 
$\phi/\phi_{c}  = 0.0732$ (lower curve) and 
$\phi/\phi_{c} = 0.988$ (upper curve). The dashed line is a fit to 
$1/f^{\alpha}$ with $\alpha = 1.0$. 
(b) The noise power $S_{0}$ vs $\phi/\phi_{c}$ 
obtained by integrating $S(\omega$) over 
a fixed interval of $\omega$.
$S_0$ peaks just below $\phi_{c}$.
}
\label{fig:5}
\end{figure}

In order to better understand the stability of the frozen states, 
we color the particles according to the net component of the motor force
along the $y$ direction.  Particles that have a net motion in the positive
$y$ direction are plotted with a dark color and a light arrow, while
particles that have a net motion in the negative $y$ direction are drawn
with a light color and a dark arrow.
Figure~\ref{fig:4}(a) shows the configuration at $\phi = 0.545$ 
in the fluctuating state 
where the system forms transient living crystals.  
A stable cluster can form  
when groups of particles come together that have
on average opposite swimming directions.
In Fig.~\ref{fig:4}(b) we illustrate
the dynamically frozen steady state at $\phi = 0.545$. 
Particles in 
the lower half of the cluster are generally moving
in the positive $y$ direction, while particles in 
the upper half of the cluster are generally moving
in the negative $y$-direction, 
so that the cluster can be regarded as a blocked or jammed state  
in which the particles can no longer move past one another.    
Figure~\ref{fig:4}(c) 
shows the dynamically frozen steady state at $\phi = 0.73$, 
where again we find a similar phenomenon in which the
lower half of the cluster is moving against the upper half.  
As the density increases, the size of the cluster grows 
until eventually the system forms
a triangular lattice with embedded voids, as shown in 
Fig.~\ref{fig:4}(d) for the dynamically frozen 
steady state at $\phi = 0.848$. 
At these high densities the 
frozen states no longer have the same clear top/bottom asymmetry features
and the crystalline state can be
stabilized with more random up/down particle arrangements. 
Eventually at $\phi = 0.9$ the system 
would naturally crystallize on its own even in the absence of 
motor forces. 
Since the number of random configurations
that lead to a stable crystalline state
increases as $\phi = 0.9$ is approached from below,
the time $\tau$ required to reach an absorbing state decreases
with increasing $\phi$ above $\phi_c$.    

The transient times $\tau$ generally increase for increasing system size
$L$; however, 
the value of $\phi_{c}$ does not shift. The dynamically frozen 
steady states are likely 
affected by the periodic boundary conditions, which limit the total
number of possible configurations. 
In any real experiments performed with active matter, the system will 
also be bounded and have a limit on the number of possible
configurations, so we expect that 
the transition to a dynamically frozen steady state that we
find can be readily observed in experiment.

\section{Noise Fluctuations}
In the sheared colloidal system of Ref.~\cite{5}, 
two transient times were observed.
The first is the time $\tau$ required for the system 
to settle into the non-fluctuating
state on the low shear
amplitude side of transition, 
and the second is the time required
for the system to settle into 
a steady fluctuating state 
as the critical shear amplitude is approached 
from above \cite{5}. 
In our system, for $\phi < \phi_{c}$ we do not find a clear
transient signature for the settling of the fluctuations into a steady
fluctuating state;
however, we can measure the
changes in the velocity fluctuations upon approaching $\phi_c$ from below.
In Fig.~\ref{fig:5}(a) we plot 
power spectra obtained from the time series of the
velocity fluctuations, 
$S(\omega) = |\int V(t) e^{-i\omega t}dt|^{2}$
at $\phi/\phi_{c} =  0.0732$ and $\phi/\phi_{c}  = 0.988$.
Here we find that upon approaching
$\phi_{c}$ from below, the noise characteristic changes
from white at low $\phi/\phi_c$
to a $1/f^{\alpha}$ form close to the transition, with a large
increase in the noise power at low frequencies.
The solid line in Fig.~\ref{fig:5}(a) is a
fit with $\alpha = 1.0$. 

\begin{figure*}
\centering
\includegraphics[width=0.7\textwidth]{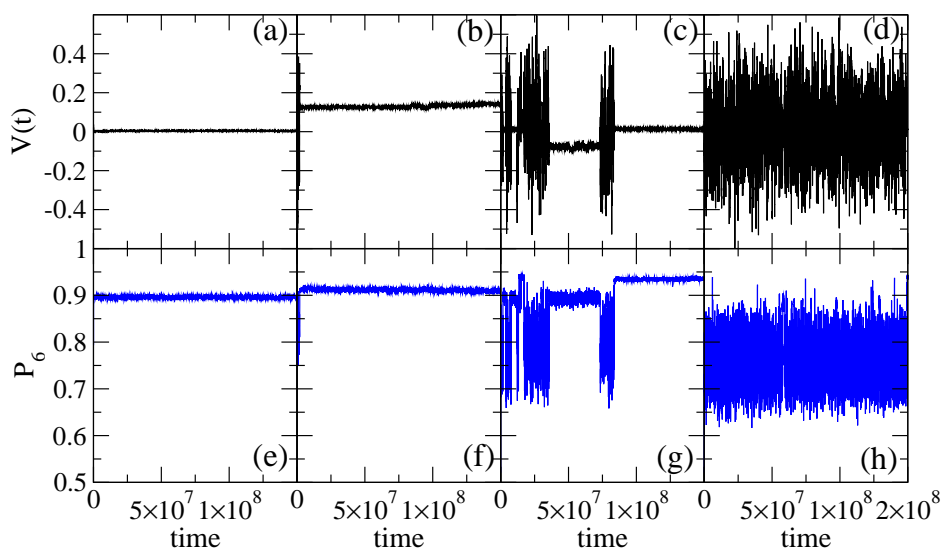}
\caption{
(a,b,c,d) $V(t)$ for a system at $\phi/\phi_{c} = 1.43$ 
where stochastic noise is added in the
form of thermal kicks of magnitude $F^T= 0$ (a), 2.0 (b), 6.0 (c), and 
9.0 (d). (e,f,g,h)
The corresponding fraction of particles
with six-fold neighbors $P_6(t)$ for a system at
$\phi/\phi_c=1.43$ and $F^T= 0$ (e), 2.0 (f), 6.0 (g), and 9.0 (h).
}
\label{fig:6}
\end{figure*}

We can further characterize the noise 
by measuring the noise power $S_0$, obtained by integrating
$S(\omega)$ over a range of low frequencies
\cite{noise,noiseA,noise2}.
In Fig.~\ref{fig:5}(b) we plot $S_{0}$ versus $\phi/\phi_{c}$ 
showing that the noise power peaks 
just below $\phi_{c}$. 
The noise measurements taken in the frozen steady state for
$\phi > \phi_{c}$ show very low noise power due to the lack of fluctuations.
A peak in the noise power has previously been proposed to indicate the 
presence of 
a critical point in both 
equilibrium \cite{50} and nonequilibrium systems \cite{51}. 
The increase in the low frequency noise power
indicates that the fluctuations are occurring on larger length scales and thus
take longer time scales to occur.  This is
correlated with the living crystals becoming
longer lived as $\phi_{c}$ is approached. 
Once the living crystal structure becomes large enough,
the system can find a dynamically frozen configuration
and then remains trapped in that state.  

\section{Thermal Noise}
We next consider the robustness of
the transition into the dynamically frozen state 
in the presence of stochastic noise, introduced by adding
Langevin kicks ${\bf F}^T$ to the equations of motion with the 
following properties: 
$\langle {\bf F}^{T}(t)\rangle = 0$
and
$\langle F^{T}_{i}(t)F_j^T(t^{\prime})\rangle = 2\eta k_BT \delta_{ij}\delta(t - t^{\prime})$. 
In the absence of other particles or a motor force, 
these kicks induce a random motion of an individual particle. 
We find that close-packed crystals at $\phi=0.9$ with no motor force
thermally disorder when
$F^T > 8.0$. 
In general we find that for $\phi > \phi_{c}$, 
a transition to the
dynamically frozen state still occurs 
provided that the thermal fluctuations are small enough; however,
even at higher values of 
$F^{T}$ we observe a distinct difference in the dynamics of the system
for $\phi > \phi_{c}$ compared to
$\phi < \phi_{c}$.

In Fig.~\ref{fig:6}(a-d) we plot $V(t)$ 
for a system at $\phi/\phi_{c} = 1.43$ 
with different thermal noise levels $F^{T} = 0.0$, 2.0, 6.0, and $9.0$. 
In Fig.~\ref{fig:6}(e-h) we show the 
corresponding fraction of six-fold
coordinated particles $P_{6}(t)$ vs time. 
To compute $P_6$, we find $z_i$, the number of neighbors of particle $i$,
from a Voronoi construction, and then 
obtain $P_6=N^{-1}\sum_i^N \delta(z_i-6)$.
In the frozen cluster state, most of the particles have six neighbors
so $P_{6} \approx 0.9$ as shown in Fig.~\ref{fig:6}(e) for $F^T=0$, where the 
velocity
fluctuations are also small as indicated in Fig.~\ref{fig:6}(a).
The transition to the quasistatic cluster state remains robust even for
finite $F^T$, as shown in Fig.~\ref{fig:6}(b,f) at $F^T=2.0$
where the system rapidly passes
from the fluctuating
state to the frozen
cluster state and remains trapped for further times with only 
minor rearrangements. 
The frozen cluster state can be regarded as a solid that only breaks or
melts once a critical level of thermal fluctuations is
added.
For increasing $F^{T}$, the behavior becomes increasingly intermittent.
A frozen cluster forms for a period of time but can undergo an instability that
sends the system back into a fluctuating state for another period of time
before a new frozen cluster appears.  This process is illustrated in
Fig.~\ref{fig:6}(c,g) for $F^T = 6.0$, where the regions of strongly 
fluctuating $V(t)$ are correlated with
strongly fluctuating values of $P_{6}(t)$. 
We do not observe this type of two-step intermittent behavior 
in the fluctuating regime
at $\phi < \phi_c$ for $F^T=0$.  The higher temperature systems no
longer become permanently absorbed into the frozen state; instead, we
observe temperature-induced transitions for $\phi>\phi_c$ from the
frozen state at low temperature to the intermittent state at 
intermediate temperatures. 
When the stochastic fluctuations are large enough the system always remains
in the fluctuating state,
as shown in Fig.~\ref{fig:6}(d,h) for $F^T=9.0$.
The value of $F^T$ at which the system remains permanently in a fluctuating
state decreases with decreasing $\phi/\phi_c$.
These results show that the transition to the dynamically frozen state can
still be realized in the presence of fluctuations
provided the
additional stochastic fluctuations are sufficiently small.

\section{Dynamic Freezing in the Presence of Obstacles}

\begin{figure}
\centering
\includegraphics[width=0.5\textwidth]{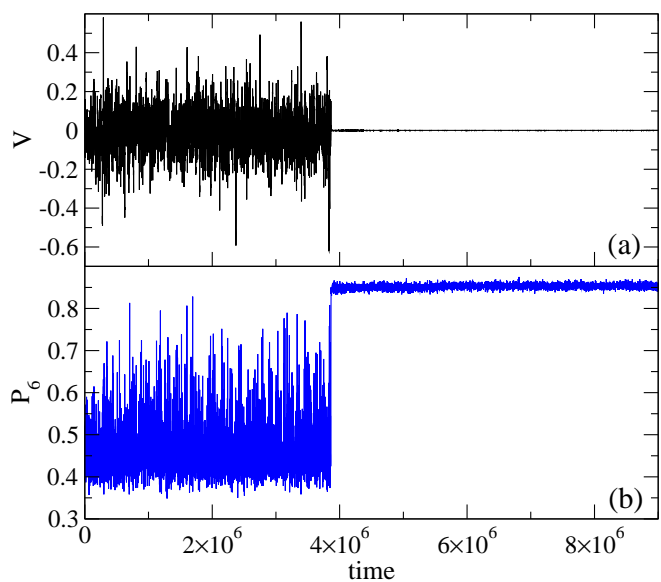}
\caption{
(a) $V(t)$ for a system at $\phi = 0.363$ 
containing $N_p=2$ obstacles or pinned disks.
(b) The corresponding $P_6(t)$.
Even though $\phi<\phi_c$,
the system shows a transition to a dynamically frozen state consisting
of a pinned cluster.   
}
\label{fig:7}
\end{figure}

We next consider the effects of adding a small number of obstacles to the system in form of
stationary or pinned disks. 
In this case we find that even a very low obstacle density can
cause the system to organize into a dynamic frozen 
state at values of $\phi$ well below the
obstacle-free critical value $\phi_{c}$; however, the nature of the frozen
states in the presence and absence of obstacles
are quite different.
Figure~\ref{fig:7}(a) shows 
$V(t)$ for a system at $\phi = 0.363$ containing $N_p=2$ obstacles.
There is a 
clear transition from a 
fluctuating state to a dynamically frozen state even though $\phi<\phi_c$.
The corresponding $P_6(t)$ in Fig.~\ref{fig:7}(b)
shows large variations in the fluctuating state with values
as large as $P_6=0.81$ indicating transient clustering, while
at the transition to the frozen state we find $P_6 \approx 0.84$
as the system forms a pinned hexagonal-faceted crystal
that is centered near one of the obstacles. 

\begin{figure}
\centering
\includegraphics[width=0.5\textwidth]{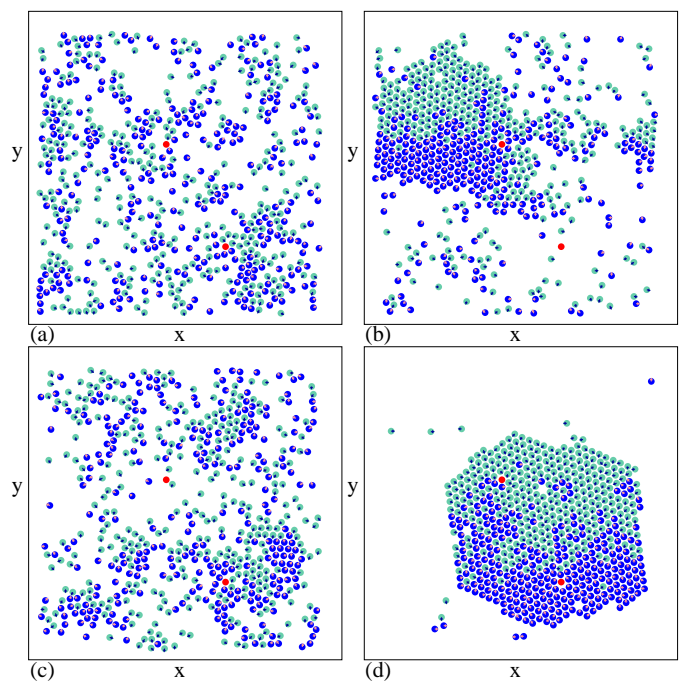}
\caption{
Snapshots of disk and obstacle positions in a system with $\phi=0.363$
and $N_p=2$
where the arrows indicate the direction of the
motor force for each disk.  
Moving disks are colored as in Fig.~\ref{fig:4}.
Red disks are immobile obstacles.
(a) The initial fluctuating state containing
small transient clusters. 
(b) At a later time, transient clusters are nucleated by the obstacles.
(c) At a still later time, the transient cluster shown in (b) has broken apart. 
(d) The dynamically frozen steady state where a faceted crystal forms 
around the obstacles.
}
\label{fig:8}
\end{figure}

The obstacles serve as cluster nucleation sites.
Figure~\ref{fig:8} shows snapshots at different times of the
disk and obstacle positions for the system in Fig.~\ref{fig:7}. 
In the initial fluctuating state in Fig.~\ref{fig:8}(a),
there are small transient clusters.
Figure~\ref{fig:8}(b) shows the system at a later time but still in 
the fluctuating
state where a cluster nucleates around one of the obstacles.
This cluster eventually breaks apart
as illustrated in Fig.~\ref{fig:8}(c) which shows an even later
time in the fluctuating state.
Figure~\ref{fig:8}(d) shows the configuration in the dynamically 
frozen steady state,
where the particles form a single cluster 
that has the shape of a faceted hexagonal crystal. 
Unlike the frozen steady state in the obstacle-free system, 
the cluster in Fig.~\ref{fig:8}(d) is not drifting but is 
pinned by the obstacles. 
The cluster exhibits the additional feature that the particles in the 
upper half of the cluster are moving in the negative $y$ direction on
average while the particles in the lower half of the cluster are moving
in the positive $y$ direction on average.
In particular, the particles on the outer edges of the crystal have their 
motor forces oriented toward the center of the cluster.
There are a small number of particles outside of the cluster 
that are moving in closed orbits. 

\begin{figure}
\centering
\includegraphics[width=0.5\textwidth]{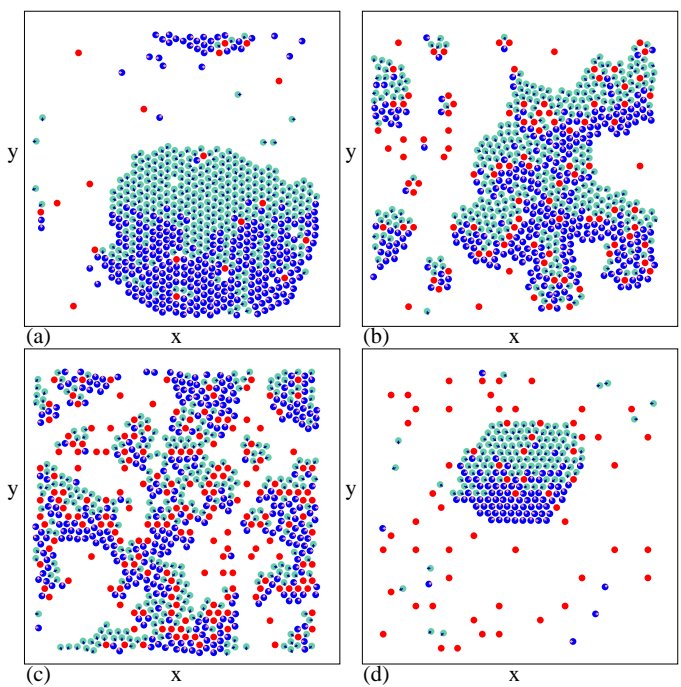}
\caption{
Snapshots of disk and obstacle positions in 
dynamically frozen states of systems containing obstacles
where the arrows indicate the direction of the motor force for each disk.  
Moving disks are colored as in Fig.~\ref{fig:4}.  Red disks are immobile
obstacles.
(a) At $\phi = 0.363$ and $N_p=20$, the cluster no longer has facets.
(b) At $\phi = 0.363$ and $N_p=100$,
the cluster starts to become much more disordered.   
(c) At $\phi = 0.363$ and $N_p=200$, multiple clusters start to form. 
(d) At $\phi = 0.1212$ and $N_p=60$ cluster formation still occurs.
}
\label{fig:9}
\end{figure}

In general for $\phi < \phi_{c}$, when a small number of obstacles 
are present the system
eventually forms a faceted pinned crystal centered around 
the defects. When more obstacles are added, 
the transient time
to reach the pinned state decreases and the 
system may form multiple pinned clusters. 
In Fig.~\ref{fig:9}(a), a system with
$\phi = 0.363$ and $N_p=20$ forms a single cluster that no longer has
the faceted features found for lower obstacle densities.
Adding more obstacles to the same system produces the frozen
steady state shown in Fig.~\ref{fig:9}(b), 
where the clustering is much more disordered when 
$N_p=100$.
Figure~\ref{fig:9}(c) shows the same system
with $N_p=200$, where multiple pinned clusters form. 
The
cluster states can also occur at lower 
densities, as illustrated in Fig.~\ref{fig:9}(d) for a system with
$\phi = 0.1212$ and $N_p=60$.

\begin{figure}
\centering
\includegraphics[width=0.5\textwidth]{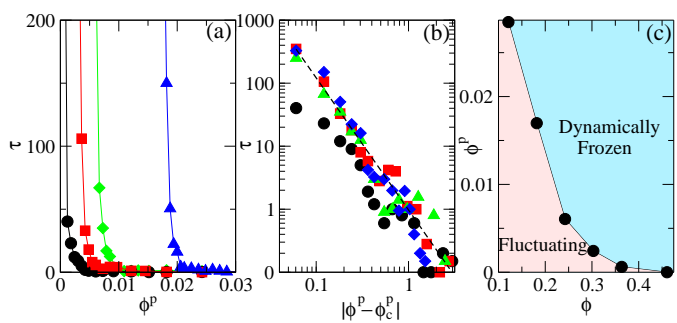}
\caption
{
(a) The transient time $\tau$ to reach a pinned cluster state 
vs the obstacle density $\phi^{p}$
for $\phi = 0.363$ (circles), $0.303$ (squares), 
$0.2424$ (diamonds), and $0.182$ (triangles).  
(b) A log-log plot of $\tau$ vs 
$|\phi^{p} - \phi^{p}_{c}|$, 
where the solid line is a fit to 
$\tau \propto |\phi^p-\phi^p_c|^{-\beta}$
with $\beta = 1.9$.    
(c) $\phi_{p}$ vs $\phi$ showing the separation between 
the region where the system remains fluctuating
and the dynamically frozen region where the
system eventually forms a pinned cluster state. 
}
\label{fig:10}
\end{figure}

For $\phi < \phi_{c}$ we 
find that there is a critical obstacle density $\phi^{p}_{c}$  below which the
system 
never reaches a dynamically frozen steady state. 
In Fig.~\ref{fig:10}(a) we plot the transient time 
$\tau$ required to reach 
a pinned cluster versus the obstacle density $\phi^{p}$ 
for $\phi = 0.363$, $0.303$, $0.2424$, and $0.182$, 
showing that
$\tau$ diverges at differing critical pinning densities
$\phi^p_c$. 
The curves can be fit to the form
$\tau \propto |(\phi^{p} - \phi^{p}_{c})|^{-\beta} $ 
with $\beta \approx 1.9 \pm 0.4$, as shown 
in the log-log plot of Fig.~\ref{fig:10}(b).
The fact that the transition
from the fluctuating to the dynamical state 
shows a different scaling than 
in the obstacle-free case 
suggests that  the addition of obstacles changes the class of the
transition.
In Fig.~\ref{fig:10}(c) 
we show the line dividing the dynamically frozen and fluctuating steady
states on a plot of obstacle density $\phi^{p}$ 
versus disk density $\phi$. 
As $\phi$ decreases, the density of obstacles
required to induce the formation of a dynamically frozen state increases.

There have been several studies examining the effect of quenched disorder on 
active matter systems.
For flocking models it was found that the coherence of the system 
passes through a maximum when 
random noise is added \cite{52,53}. 
For run and tumble dynamics it was shown that the effective mobility decreases 
with increasing run length \cite{54}. 
In both classes of system, 
a fluctuating state is always present, 
so there is not a sharp transition to a completely pinned state
as we observe in this work.

\section{Summary}
We examine a system of run-and-tumble disks 
in the limit of infinite run time and find 
that above a critical density, a transition occurs
from a fluctuating state to an absorbing quiescent state. 
In the dynamically frozen state, 
a drifting cluster forms and the fluctuations 
in the system almost completely vanish. 
The transient time required for the system to organize 
into the dynamically frozen state diverges as a power law as the critical
density is approached with 
an exponent of $\nu_{||} \approx 1.21$. 
This behavior is very similar to that observed
in the random organization transition 
to a non-fluctuating state in sheared colloidal systems 
where the transient times
also show power law divergences with comparable critical exponents. 
A key difference between the
random organization and active matter systems 
is that in the quiescent state, the colloidal particles cease to
interact with each other, while
the active matter particles form a strongly correlated or jammed configuration
in which most of the particles are in constant contact.
We also find that the magnitude of the fluctuations 
diverge as $\phi_c$ is approached from below.
The dynamic transition is robust against the addition of
a certain range of stochastic or thermal fluctuations.    
When a small number of obstacles are added to the system, 
the transition to a frozen state can occur at
densities much lower than the obstacle-free 
critical density, and in this case the frozen state is 
characterized by the formation of pinned faceted crystals. 

\section{Acknowledgements}
This work was carried out under the auspices of the 
NNSA of the 
U.S. DoE
at 
LANL
under Contract No.
DE-AC52-06NA25396.



\footnotesize{

\end{document}